\documentclass[eqsecnum,twocolumn,aps,epsf,showpacs]{revtex4}
\usepackage{graphicx}
\usepackage{color}
\begin{document}
\draft

\title{Cu-spin dynamics in the overdoped regime of La$_{2-x}$Sr$_x$Cu$_{1-y}$Zn$_y$O$_4$ probed by muon spin relaxation}

\author{Risdiana}
\altaffiliation[Present address: ]{Department of Physics, Faculty of Mathematics and Natural Sciences, Padjadjaran University, Jl. Raya Bandung-Sumedang Km. 21 Jatinangor, West Jawa, Indonesia 45363.}
\author{T. Adachi}
\thanks{Corresponding author: adachi@teion.apph.tohoku.ac.jp}
\author{N. Oki}
\author{S. Yairi}
\author{Y. Tanabe}
\author{K. Omori}
\author{Y. Koike}
\affiliation{Department of Applied Physics, Graduate School of Engineering, Tohoku University, 6-6-05 Aoba, Aramaki, Aoba-ku, Sendai 980-8579, Japan}

\author{T. Suzuki}
\author{I. Watanabe}
\affiliation{Advanced Meson Science Laboratory, The Institute of Physical and Chemical Research (RIKEN), 2-1 Hirosawa, Wako 351-0198, Japan}

\author{A. Koda}
\affiliation{Muon Science Laboratory, Institute of Materials Structure Science, High Energy Accelerator Research Organization (KEK-IMSS), 1-1 Oho, Tsukuba 305-0801, Japan}

\author{W. Higemoto}
\affiliation{Advanced Science Research Center, Japan Atomic Energy Agency (JAEA), 2-4 Shirane, Shirakata, Tokai, Naka, Ibaraki 319-1195, Japan}

\date{\today}

\begin{abstract}
Muon-spin-relaxation measurements have been performed for the partially Zn-substituted La$_{2-x}$Sr$_x$Cu$_{1-y}$Zn$_y$O$_4$ with $y = 0 - 0.10$ in the overdoped regime up to $x = 0.30$. 
In the 3 \% Zn-substituted samples up to $x = 0.27$, exponential-like depolarization of muon spins has been observed at low temperatures, indicating Zn-induced slowing-down of the Cu-spin fluctuations. 
The depolarization rate decreases with increasing $x$ and almost no fast depolarization of muon spins has been observed for $x = 0.30$ where superconductivity disappears. 
The present results suggest that the dynamical stripe correlations exist in the whole superconducting regime of La$_{2-x}$Sr$_x$CuO$_4$ and that there is no quantum critical point at $x \sim 0.19$.

\end{abstract}
\vspace*{2em}
\pacs{74.62.Dh, 74.72.Dn, 76.75.+i, 74.25 Ha}
\maketitle
\newpage

\section{Introduction}\label{intro}
Effects of nonmagnetic impurities on the Cu-spin dynamics have been one of central issues in the research of high-$T_{\rm c}$ superconductivity. 
It is widely recognized that local magnetic moments due to Cu spins are induced by the nonmagnetic Zn in the underdoped regime~\cite{mahajan} and that the superconductivity is strongly suppressed by Zn.~\cite{maeno}
In the overdoped regime, on the other hand, Panagopoulos {\it et al}. have performed muon-spin-relaxation ($\mu$SR) measurements in the Zn-substituted La$_{2-x}$Sr$_x$Cu$_{1-y}$Zn$_y$O$_4$ with $y=0.05$ and have found Zn-induced slowing-down of the Cu-spin fluctuations for $x<0.19$ while not for $x \ge 0.19$, pointing out the existence of a quantum critical point (QCP) at $x \sim 0.19$.~\cite{panago} 

Dynamical stripe correlations of spins and holes have attracted great interest in relation to the mechanism of high-$T_{\rm c}$ superconductivity.~\cite{tra1,tra2} 
It is well known that the dynamical stripe correlations tend to be pinned and statically stabilized not only by the characteristic structure of the tetragonal low-temperature structure (space group: $P4_2/ncm$) in the La-214 system but also by Zn impurities.~\cite{koike1,koike2,adachi1,smith} 
So far, impurity effects on the Cu-spin dynamics and superconductivity have been investigated from the zero-field (ZF) $\mu$SR measurements in the underdoped regime of La$_{2-x}$Sr$_x$Cu$_{1-y}$Zn$_y$O$_4$ with $x \le 0.15$.~\cite{nabe1,nabe2,nabe3,adachi2,adachi3,adachi4,koike3} 
It has been found that fluctuations of Cu spins around Zn exhibit slowing down due to the pinning of the dynamical stripes by Zn, leading to the formation of a static stripe order and the destruction of superconductivity around Zn. 
This is an interpretation called a stripe-pinning model. 
Similar effects of Zn on the Cu-spin fluctuations have been observed in the underdoped regime of Bi$_2$Sr$_2$Ca$_{1-x}$Y$_x$(Cu$_{1-y}$Zn$_y$)$_2$O$_{8+\delta}$~\cite{nabe-bi1,nabe-bi2} and YBa$_2$Cu$_{3-2y}$Zn$_{2y}$O$_{7-\delta}$~\cite{ako} as well. 
The static stripe order is regarded as competing with the superconductivity. 
On the other hand, the dynamical stripe correlations may play an important role in the appearance of superconductivity. 
In fact, theoretical models on the mechanism of superconductivity based upon the dynamical stripe correlations have been proposed.~\cite{theo} 
If this is the case, similar effects of Zn must be observed not only in the underdoped regime but also in the overdoped regime where superconductivity appears.

In this paper, we investigate effects of Zn on the Cu-spin dynamics from the $\mu$SR measurements in the overdoped regime of La$_{2-x}$Sr$_x$Cu$_{1-y}$Zn$_y$O$_4$ with $0.15 \le x \le 0.30$ and $0 \le y \le 0.10$, in order to elucidate whether the stripe-pinning model holds good even in the overdoped regime of La$_{2-x}$Sr$_x$CuO$_4$ or not.

\section{Experimental}
\begin{figure}[tbp]
\begin{center}
\includegraphics[width=0.6\linewidth]{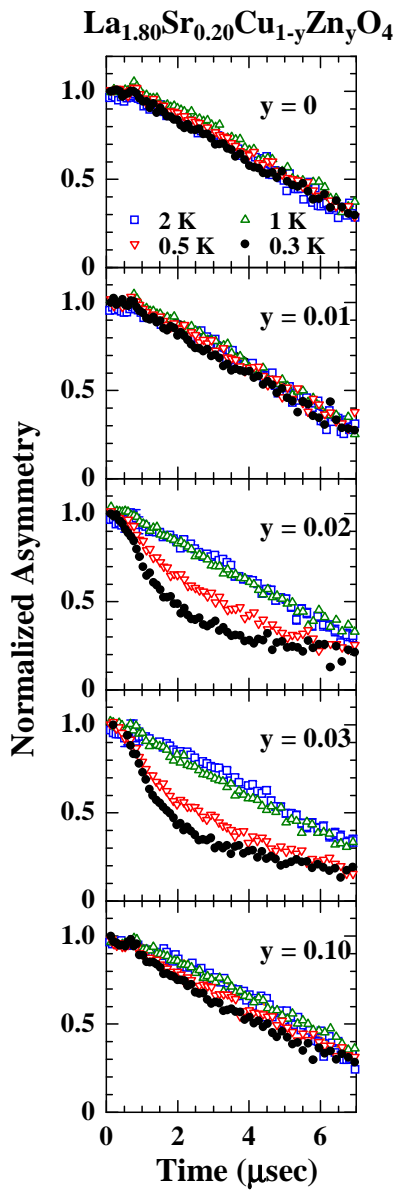}
\end{center}
\caption{(color online) ZF-$\mu$SR time spectra of La$_{1.80}$Sr$_{0.20}$Cu$_{1-y}$Zn$_y$O$_4$ with various $y$ values at low temperatures down to 0.3 K.}  
\label{fig:fig1} 
\end{figure}

Polycrystalline samples of La$_{2-x}$Sr$_x$Cu$_{1-y}$Zn$_y$O$_4$ with $0.15 \le x \le 0.30$ and $0 \le y \le 0.10$ were prepared by the ordinary solid-state reaction method. 
Raw materials of dried La$_2$O$_3$, SrCO$_3$, CuO and ZnO powders were mixed in a stoichiometric ratio and prefired in air at 750 $^{\rm o}$C for 3 h, followed by prefiring at 900 $^{\rm o}$C for 12 h. 
The prefired materials were reground and pressed into pellets of 10 mm in diameter, and sintered in air at 750 $^{\rm o}$C for 3 h, followed by sintering at 1050 $^{\rm o}$C for 24 h with repeated regrinding. 
The as-grown samples were annealed in flowing gas of O$_2$ at 500 $^{\rm o}$C for 72 h. 
All of the samples were checked by the powder x-ray diffraction measurements to be single phase. 
Electrical resistivity measurements were also carried out to check the quality of the samples, which was found to be good. 

The ZF-$\mu$SR measurements were performed at low temperatures down to 0.3 K at the RIKEN-RAL Muon Facility at the Rutherford-Appleton Laboratory in the UK and at lower temperatures down to 0.02 K at the KEK-MSL in Japan, using a pulsed positive surface muon beam. 
The asymmetry parameter $A(t)$ at a time $t$ was defined as $A(t) = [F(t) - \alpha B(t)] / [F(t) +  \alpha B(t)]$, where $F(t)$ and $B(t)$ were total muon events counted by the forward and backward counters, respectively, and $\alpha$ is the calibration factor reflecting the relative counting efficiencies between the forward and backward counters.

\section{Results}
It is known that muons with positive charges injected in a high-$T_{\rm c}$ copper oxide tend to stop near oxygen ions.~\cite{muon1} 
In La$_{2-x}$Sr$_x$CuO$_4$, it is well established that muons tend to stop only near the so-called apical oxygen just above Cu.~\cite{muon2} 
Therefore, muons stopping near the apical oxygen feel strong internal fields from the CuO$_2$ plane, when the Cu-spin fluctuations exhibit slowing down at low temperatures.

Figure 1 shows the ZF-$\mu$SR time spectra of La$_{2-x}$Sr$_x$Cu$_{1-y}$Zn$_y$O$_4$ with $x = 0.20$ and $0 \le y \le 0.10$. 
At high temperatures above 1 K, all the spectra show Gaussian-like depolarization due to dipole fields induced by randomly oriented nuclear spins, indicating a fast fluctuating state of Cu spins beyond the $\mu$SR time window ($10^{-6} - 10^{-11}$ s).  
At 0.3 K, the Gaussian-like depolarization is still observed for $y = 0$ and 0.01. 
On the other hand, the muon-spin depolarization becomes fast for $y = 0.02$ and 0.03 so that exponential-like depolarization is observed. 
This indicates slowing down of the Cu-spin fluctuations, as in the case of the underdoped regime.~\cite{nabe1,nabe2,nabe3,adachi2,adachi3,adachi4,koike3} 
Even at 0.3 K, however, no muon-spin precession corresponding to the formation of any coherent magnetic order is observed. 
In the heavily Zn-substituted sample with $y = 0.10$, no fast depolarization of muon spins is observed, indicating that the Cu spins again fluctuate fast beyond the $\mu$SR time window. 
This is explained as being due to the spin dilution through the large amount of substitution of nonmagnetic Zn, namely, due to the destruction of the spin correlation, as in the case of the underdoped regime.~\cite{nabe1,nabe2,nabe3,adachi2,adachi3,adachi4,koike3} 

\begin{figure}[tbp]
\begin{center}
\includegraphics[width=1.0\linewidth]{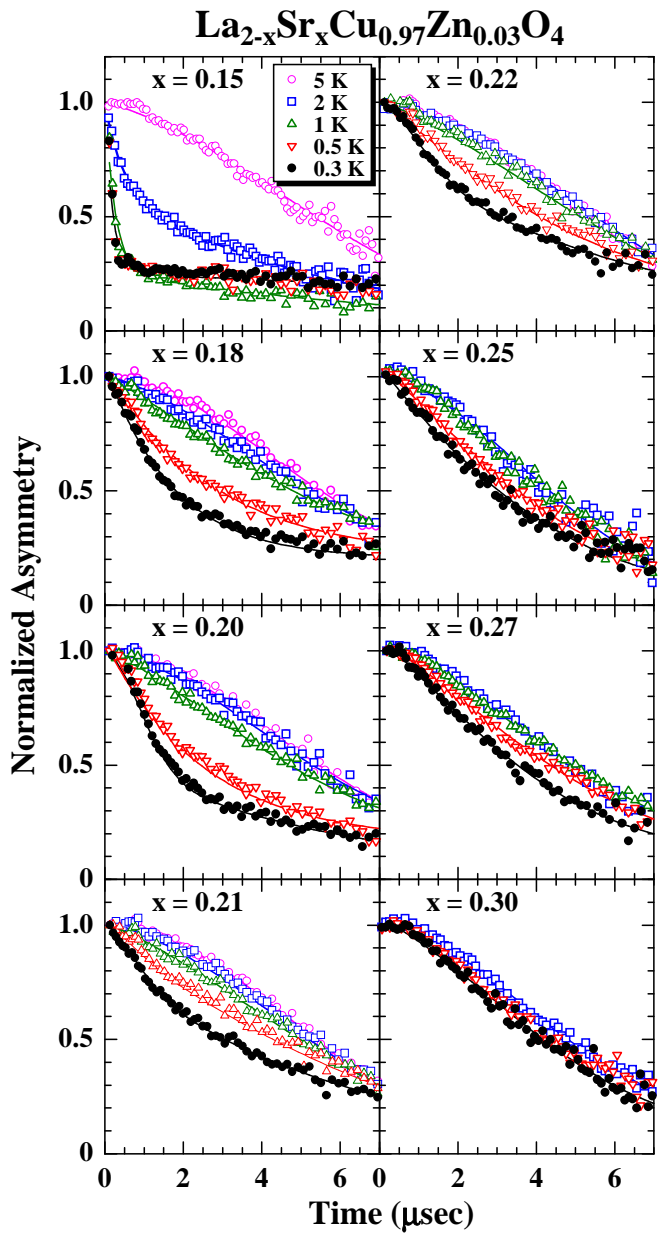}
\end{center}
\caption{(color online) ZF-$\mu$SR time spectra of La$_{2-x}$Sr$_x$Cu$_{0.97}$Zn$_{0.03}$O$_4$ with $0.15 \le x \le 0.30$ at low temperatures down to 0.3 K. Solid lines indicate the best-fit results using $A(t) = A_0 e^{- \lambda_0 t} G_{\rm Z}(\Delta, t) + A_1 e^{- \lambda_1 t}$.}  
\label{fig:fig2} 
\end{figure}

Figure 2 shows the ZF-$\mu$SR time spectra of the 3 \% Zn-substituted La$_{2-x}$Sr$_x$Cu$_{1-y}$Zn$_y$O$_4$ with $0.15 \le x \le 0.30$ and $y = 0.03$ at low temperatures down to 0.3 K. 
The reason why the 3 \% Zn-substituted samples are taken is that the Zn-induced slowing-down of the Cu-spin fluctuations is most observable in the overdoped regime as well as in the underdoped regime.~\cite{nabe1,nabe2,nabe3,adachi2,adachi3,adachi4,koike3} 
Focusing on the spectra at 0.3 K, the exponential-like depolarization is still observed with increasing $x$ up to $x = 0.27$, indicating the slowing down of the Cu-spin fluctuations. 
The depolarization becomes weak with increasing $x$ and almost no fast depolarization of muon spins is observed for $x = 0.30$ where superconductivity disappears. 
As for $x = 0.15$, both the very fast depolarization in a short-time region below 1 $\mu$s and the almost flat spectrum above 1 $\mu$s suggests the formation of a static magnetic order, as in the underdoped La$_{2-x}$Sr$_x$Cu$_{1-y}$Zn$_y$O$_4$ around $x = 0.115$.~\cite{nabe1,nabe2,nabe3,adachi2,adachi3} 
It is noted that similar spectra have been observed in the underdoped La$_{2-x}$Sr$_x$CuO$_4$ and Y$_{0.8}$Ca$_{0.2}$Ba$_2$Cu$_3$O$_6$ with $p$ (the hole concentration per Cu) $=0.09$~\cite{nieder2} and (Ca$_x$La$_{1-x}$)(Ba$_{1.75-x}$La$_{0.25+x}$)Cu$_3$O$_y$.~\cite{kanigel}
These spectra suggest that an incoherent magnetic order is formed due to the spin correlation not being so long-ranged. 

In order to obtain detailed information on the Cu-spin dynamics, the time spectra were analyzed using the following two-component function:
\begin{equation}
A(t) = A_0 e^{- \lambda_0 t} G_{\rm Z}(\Delta, t) + A_1 e^{- \lambda_1 t}.
\label{eq1}
\end{equation}
The first term represents the slowly depolarizing component in a region where the Cu spins fluctuate fast beyond the $\mu$SR time window. 
The $A_0$ and $\lambda_0$ are the initial asymmetry and the depolarization rate of the slowly depolarizing component, respectively. 
The $G_{\rm Z}(\Delta ,t)$ is the static Kubo-Toyabe function with a distribution width, $\Delta$, of nuclear-dipole fields at the muon site.~\cite{muon3} 
The second term represents the fast depolarizing component in a region where the Cu-spin fluctuations slow down. 
The $A_1$ and $\lambda_1$ are the initial asymmetry and the depolarization rate of the fast depolarizing component, respectively. 
The time spectra are well fitted with Eq. (\ref{eq1}), as shown by solid lines in Fig. 2. 
In the fitting, the value of $A_0 + A_1$ was kept constant at all measured temperatures. 

\begin{figure}[tbp]
\begin{center}
\includegraphics[width=0.8\linewidth]{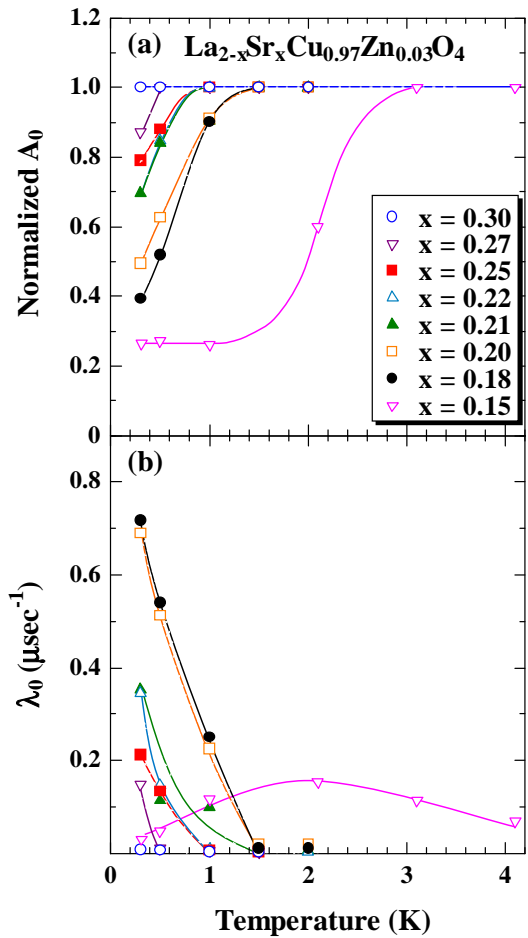}
\end{center}
\caption{(color online) (a) Temperature dependence of the initial asymmetry of the slowly depolarizing component, $A_0$, for La$_{2-x}$Sr$_x$Cu$_{0.97}$Zn$_{0.03}$O$_4$ with $0.15 \le x \le 0.30$. Values of $A_0$ are normalized by those obtained at high temperatures. Lines are to guide the reader's eye. (b) Temperature dependence of the depolarization rate of the slowly depolarizing component, $\lambda_0$, for La$_{2-x}$Sr$_x$Cu$_{0.97}$Zn$_{0.03}$O$_4$ with $0.15 \le x \le 0.30$. Lines are to guide the reader's eye.}  
\label{fig:fig3} 
\end{figure}

Figure 3(a) shows the temperature dependence of $A_0$ for the 3 \% Zn-substituted La$_{2-x}$Sr$_x$Cu$_{0.97}$Zn$_{0.03}$O$_4$ with various $x$ values. 
Values of $A_0$ are normalized by those obtained at high temperatures. 
It is found that $A_0$ decreases with decreasing temperature at low temperatures for $0.15 \le x \le 0.27$. 
The temperature where $A_0$ starts to decrease decreases with increasing $x$ and the decrease in $A_0$ disappears at $x = 0.30$. 
The decrease in $A_0$ means the appearance of the fast depolarizing component due to the slowing down of the Cu-spin fluctuations. 
As for $x = 0.15$, the $A_0$ value is saturated at low temperatures to be roughly 1/3, suggesting the formation of a static magnetic order, namely, the static stripe order.~\cite{nabe1,nabe2,nabe3,adachi2,adachi3} 

Figure 3(b) shows the temperature dependence of $\lambda_0$ for the 3 \% Zn-substituted La$_{2-x}$Sr$_x$Cu$_{0.97}$Zn$_{0.03}$O$_4$ with various $x$ values. 
Values of $\lambda_0$ increase with decreasing temperature at low temperatures for $0.18 \le x \le 0.27$. 
It is found that $\lambda_0$ at the lowest temperature of 0.3 K decreases with increasing $x$, suggesting that the degree of the slowing down of the Cu-spin fluctuations is weakened with increasing $x$. 
As for $x = 0.15$, $\lambda_0$ exhibits a peak at 2 K where $A_0$ rapidly decreases. 
This is a typical behavior of $\lambda_0$ in the case that a static magnetic order is formed.~\cite{nabe2} 

\section{Discussion}
\begin{figure}[tbp]
\begin{center}
\includegraphics[width=0.8\linewidth]{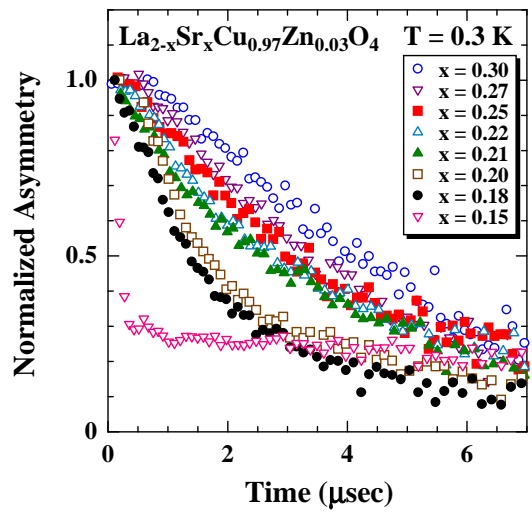}
\end{center}
\caption{(color online) ZF-$\mu$SR time spectra of La$_{2-x}$Sr$_x$Cu$_{0.97}$Zn$_{0.03}$O$_4$ with $0.15 \le x \le 0.30$ at 0.3 K.}  
\label{fig:fig4} 
\end{figure}

The most important feature in the present results is that the fast muon-spin depolarization due to the Zn-induced slowing-down of the Cu-spin fluctuations is observed in the overdoped regime where superconductivity appears, though the muon-spin depolarization becomes weak with increasing $x$. 
This is clearly seen in Fig. 4. 
\begin{figure}[tbp]
\begin{center}
\includegraphics[width=0.6\linewidth]{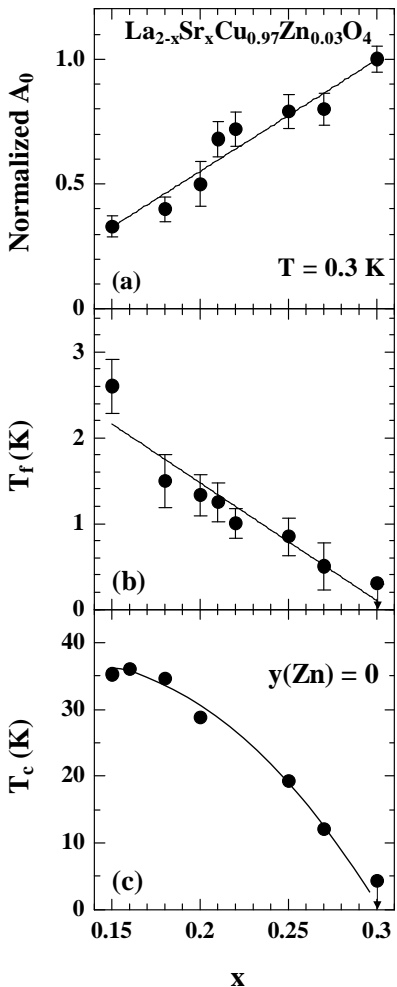}
\end{center}
\caption{(a) The initial asymmetry of the slowly depolarizing component, $A_0$, for La$_{2-x}$Sr$_x$Cu$_{0.97}$Zn$_{0.03}$O$_4$ with $0.15 \le x \le 0.30$ at 0.3 K. Values of $A_0$ are normalized by those obtained at high temperatures, (b) $T_{\rm f}$, defined as the temperature where the ZF-$\mu$SR time spectrum starts to deviate from the Gaussian-like to exponential-like behavior for La$_{2-x}$Sr$_x$Cu$_{0.97}$Zn$_{0.03}$O$_4$ with $0.15 \le x \le 0.30$. (c) The superconducting transition temperature, $T_{\rm c}$, defined at the midpoint of the resistive transition for the Zn-free La$_{2-x}$Sr$_x$CuO$_4$.~\cite{koike4}}  
\label{fig:fig5} 
\end{figure}
Quantitatively speaking, the normalized value of $A_0$ increases little by little with increasing $x$, as shown in Fig. 5(a). 
It reaches the unity at $x = 0.30$, indicating that the Cu spins fluctuate very fast with shorter periods than the $\mu$SR time window. 
The superconductivity disappears at $x \ge 0.30$,~\cite{koike4} as shown in Fig. 5(c). 
These results strongly suggest that the stripe-pinning model holds good even in the overdoped regime. 
That is, it is likely that the dynamical stripe correlations exist in the whole superconducting regime and disappear in the normal-state regime of $x \ge 0.30$. 
Therefore, it is possible that the dynamical stripe correlations play an important role in the appearance of superconductivity. 
The present results are supported by the low-energy inelastic neutron-scattering experiment in the overdoped regime of La$_{2-x}$Sr$_x$CuO$_4$ by Wakimoto {\it et al}.~\cite{wakimoto} 
They have reported that the maximum value of the integrated dynamic spin susceptibility of the incommensurate magnetic peaks corresponding to the dynamical stripe correlations decreases linearly with increasing $x$ for $x \ge 0.25$ and disappears at $x = 0.30$. 
It is noted that the existence of Zn-induced local moments~\cite{mahajan} does not conflict with the stripe-pinning model. 
It is because, even in the stripe-ordered state pinned by Zn, the spin correlation is locally destroyed in the nearest neighborhood of Zn, producing local magnetic moments around Zn.~\cite{local} 

The existence of QCP in the phase diagram of the temperature vs. $p$ for the high-$T_{\rm c}$ superconductors is one of interesting issues in recent years.~\cite{tallon,uchida} 
Panagopoulos {\it et al}.~\cite{panago} have reported the existence of QCP at $x \sim 0.19$ from the $\mu$SR results that no Zn-induced slowing-down of the Cu-spin fluctuations was observed in the overdoped regime above $x = 0.19$ for the 5 \% Zn-substituted La$_{2-x}$Sr$_x$Cu$_{0.95}$Zn$_{0.05}$O$_4$. 
As mention above, however, the 5 \% substitution of Zn is too much for the slowing down of the Cu-spin fluctuations to be observed. 
In our $\mu$SR results of the 3 \% Zn-substituted samples, there is no large difference between above and below $x = 0.19$. 
Accordingly, it is concluded that no QCP exists at $x \sim 0.19$. 

Figure 5(b) shows the $x$ dependence of $T_{\rm f}$, defined as the temperature where the ZF-$\mu$SR time spectrum starts to deviate from the Gaussian-like to exponential-like behavior for the 3 \% Zn-substituted La$_{2-x}$Sr$_x$Cu$_{0.97}$Zn$_{0.03}$O$_4$ with $0.15 \le x \le 0.30$. 
The value of $T_{\rm f}$ gradually decreases with increasing $x$ and tends to be zero at $x = 0.30$. 
Therefore, it is possible that QCP is located at $x = 0.30$ rather than $x = 0.19$. 

Next, we discuss the reason for the decrease of the depolarization rate of muon spins with increasing $x$. 
There are two possible reasons. 
One is due to the weakening of the correlation between Cu spins, namely the stripe correlations on account of the increase of holes. 
As well known, a hole incorporated at an oxygen site in the CuO$_2$ plane forms a Zhang-Rice singlet together with a Cu spin, producing a spin defect. 
Therefore, the increase of holes may weaken the Cu-spin correlation in the overdoped regime, leading to the decrease of the muon-spin depolarization rate with increasing $x$. 
The other reason is due to the decrease of the superconducting volume fraction with increasing $x$ in the overdoped regime. 
First, it has been suggested from transverse-field $\mu$SR measurements in the overdoped regime of Tl$_2$Ba$_2$CuO$_{6+\delta}$, because the superconducting carrier density divided by the effective mass, $n_{\rm s} / m^*$, has been found to decrease with increasing $p$.~\cite{uemura,nieder} 
Recently, the decrease of the superconducting volume fraction with increasing $x$ has been confirmed by Tanabe {\it et al}.~\cite{tanabe1,tanabe2} and Adachi {\it et al}.~\cite{adachi5} from magnetic-susceptibility measurements in the overdoped regime of La$_{2-x}$Sr$_x$CuO$_4$. 
Accordingly, it is plausible that a phase separation into superconducting and normal-state regions takes place in the high-$T_{\rm c}$ copper oxides. 
As the Cu-spin correlation is regarded as being very weak in the normal state region, the Zn-induced slowing-down of the Cu-spin fluctuations may occur only in the superconducting region. 
Therefore, the muon-spin depolarization rate may decrease with the progressive decrease of the superconducting volume fraction with increasing $x$ in the overdoped regime. 
It is a forthcoming issue to determine which reason is plausible. 

\section{Summary}
We have investigated the Cu-spin dynamics from ZF-$\mu$SR measurements in the overdoped regime of La$_{2-x}$Sr$_x$Cu$_{1-y}$Zn$_y$O$_4$ with $0.15 \le x \le 0.30$ and $0 \le y \le 0.10$. 
In the 3 \% Zn-substituted samples with $0.15 \le x \le 0.30$ and $y = 0.03$, exponential-like depolarization of muon spins has been observed at 0.3 K, indicating Zn-induced slowing-down of the Cu-spin fluctuations. 
Almost no fast depolarization of muon spins has been observed for $x = 0.30$ where the superconductivity disappears. 
The present results suggest that the dynamical stripe correlations of spins and holes exist in a wide range of $x$ where superconductivity appears and that the stripe-pinning model holds good in the overdoped regime as well as in the underdoped regime. 
This means that it is possible that the dynamical stripe correlations play an important role in the appearance of superconductivity in the high-$T_{\rm c}$ copper oxides. 
The present results also suggest that no QCP exists at $x \sim 0.19$.  

\section*{Acknowledgments}
We would like to thank K. Hachitani and H. Sato for their technical support in the $\mu$SR measurements. 
We are also grateful to Profs. K. Nagamine and K. Nishiyama for their continuous encouragement. 
This work was partly supported by Joint Programs of the Japan Society for the Promotion of Science, by a TORAY Science and Technology Grant and by a Grant-in-Aid for Scientific Research from the Ministry of Education, Culture, Sport, Science and Technology, Japan.


\begin{references}
\bibitem{local}
In the overdoped regime, to our knowledge, there have been no reports of the existence of Zn-induced local moments to date. However, based on the present results that the Zn-induced slowing-down of the Cu-spin fluctuations was observed, it is guessed that Zn tends to induce local moments even in the overdoped regime, though the size of moments decreases with an increase of $x$.
\end{references}
\end{document}